\documentstyle[editedvolume,psfig]{crckapb}

\begin{opening}
\title{COLD GAS AT HIGH REDSHIFT}

\author{Colin A. Norman}
\institute{Dept. of Physics and Astronomy\
           Johns Hopkins University\
            and\
            Space Telescope Science Institute}
\author {Robert Braun}
\institute{Netherlands Foundation for Research in Astronomy}
\end{opening}

\runningtitle{ COLD GAS AT HIGH REDSHIFT}

\begin{document}
\centerline{\it To appear in: ``Cold Gas at High Redshift''}
\centerline{\it Eds. M.Bremer et al. (Kluwer, Dordrecht)}

\begin{abstract}

We discuss the current observational and theoretical issues concerning
cold gas at high redshift and present simulations showing how a number
of observational issues can be resolved with planned future
instrumentation.

\end{abstract}

\vskip -1cm
\section{Introduction}

The observable history of the universe is dominated by a long phase
from the epoch of recombination (at redshift 1500) to the reheating and
reionization
phase (perhaps near redshift 7) when the entire intergalactic medium
is cold neutral
gas. Current limits from QSO absorption line studies place this epoch
above redshift 5. The fluctuations in this gas are so small that it is
difficult to see either in emission or absorption (Scott and Rees
1990). However, it is an interesting scientific goal to try to observe
this cool component of the intergalactic medium at high redshift. The
only objects we know something about at the highest redshifts are the
quasars. The space density of high redshift quasars clearly exhibit a
steep rise and fall about a redshift of 2-3 (Shaver 1995) and the rise
may be associated with the onset of galaxy formation.

In adiabatic models, where massive pancake structures formed and
subsequently lumps of order the size of galaxies fragmented out of
their collapse, the atomic masses of the cool gaseous pancake
structures were estimated to be up to $\sim 10^{14} - 10^{15}
M_{\odot}$. If such masses of diffuse atomic gas existed at
z~$\sim$~3.5, they would already have been detected by current searches
(Wieringa, De Bruyn and Katgert 1992 and references therein). Their
non-detection can now be understood in light of the constraints set by
microwave background studies and related research on the fluctuation
spectrum (c.f. Scott, Silk and White 1995). The relative smoothness of
the density fluctuations and the essentially mandated bottom-up nature
of the galaxy formation process greatly limits the possibilities for
directly observing proto-cluster size fluctuations in the cool gas
phase. More ingenious methods, which probe
both smaller and larger angular scales and
in particular smaller masses, are likely to be required as discussed later.

Great hopes for this meeting lay in a number of reported observations
of molecules observed at high redshift. However, while there are still
very interesting as yet unconfirmed claims of large molecular masses of
CO at high redshift the only well confirmed CO observations seen in
emission are due to the two well known gravitational lensed objects the
Cloverleaf and FSC 10214+4724 (Barvainis 1995, Scoville et al 1995, and
Frayer 1995). There have also been several detections of CO in
absorption against background radio sources in the mm band but also
associated with lensing (Combes and Wiklind, these proceedings).

With combined Keck and HST data, remarkable progress has been made in
the study of the absorption lines of QSOs and the objects that are
associated with the absorbing material. The population of Lyman Alpha
clouds can have a number of progenitors as we shall discuss. Both the
Damped Lyman Alpha (DLA) systems and the Lyman Alpha forest lines may
account for a significant fraction of the currently observed baryonic
content of galaxies (c.f. Storrie-Lombardi et al 1995).

In this review of a very large subject we focus on a brief
observational and theoretical overview of the subject of cool gas in
the universe. In particular we present 9 figures that show how current
and planned future instrumentation can detect and image cool gas at
high redshift and indicate how such observations may help resolve some
of the key issues.

\section{Neutral Gas}

 There are now several new aspects to the study of Lyman Alpha
absorption systems (c.f. Meylan 1995). From the point of view of this
workshop it seems most interesting to emphasize that recently there has
been a significant change in ideas about the origin of the Lyman Alpha
systems. In particular they seem to have correlation scales of order
$\sim$~1~Mpc and cannot be associated directly with individual
galaxies. Structure formation can produce sheetlike debris of low
column density that can account for many of the properties of the
absorption lines. More generally, remarkable simulations presented at
this meeting indicate that the distribution of the absorbers can be
obtained in N-body/SPH simulations.

\begin{figure}[htb]
\caption{We show the limiting column densities that we can expect to
detect with current and planned instrumentation in the red-shifted
21~cm line. A background flux of 100~mJy is assumed, as well as a
hydrogen spin temperature of 1000~K and a linewidth of 10~km~s$^{-1}$.
We compare these limits with the known distribution (from Petitjean {\it
et al.\ } 1993) of the column density of absorbers derived from QSO
absorption line studies.
}
\end{figure}

 At low red-shifts, it is clear that HI emission maps going to ever
fainter column densities such as the map of M81 by Yun et al (1994) are
a most interesting compliment to the rapidly advancing knowledge we are
obtaining from Keck and HST on the low column density environments of
galaxies. There is no substitute for an unbiased spatial tracer of
column density like that of an optically thin emission line.
Unfortunately, the column density sensitivity in HI emission at a fixed
physical resolution diminishes at least as rapidly as D$_{Ang}^2$, so
that only the highest column density disks will remain accessible out to
large distances, and then only with the largest possible collecting
areas, as we will see below.

However, lower column densities of HI can still be probed out to large
distances using the HI 21~cm line in absorption. In Figure~1 we show
the 5$\sigma$ limiting column densities in the 21~cm line that we can
expect to detect with the up-graded, frequency-agile WSRT (Westerbork
Synthesis Radio Telescope) and the proposed SKAI (Square Kilometer
Array Interferometer, described by Braun in these proceedings)
in a 24 hour integration. These limits were
calculated with the conservative assumptions that only a relatively
faint background source of 100~mJy flux be available and that the mean
spin temperature of the gas be 1000~K. A brighter background source or
cooler spin temperature result in a linear improvement of the column
density limit. These detection limits are superposed on the observed
number distribution of Lyman Alpha absorbers as function of column
density (from Petitjean {\it et al.\ } 1993).

It is clear from the
figure that the WSRT will allow access to the entire distribution of
Damped Lyman Alpha systems (N$_{HI}~>~10^{20.2}$~cm$^{-2}$), while the
SKAI will also permit study of much of the column density range of the
Lyman limit systems ($10^{20.2}$ $>$ N$_{HI}$ $>$ $10^{17}$~cm$^{-2}$). The
21~cm data provide additional insight into the physical properties of
the absorbing gas via an estimate of the effective spin temperature, as
well as providing an opportunity to image absorber structure at
milli-arcsec resolution utilizing VLBI (Very Long Baseline
Interferometry). The equation relating the HI column density and the
21~cm line opacity is:

\begin{eqnarray}
N_{HI} & = & {32 \pi k \nu_{21}^2 \over 3 h c^3 A_{21}} \ T_S \int \tau dV \\
       & = & 1.83 \times 10^{18} \ T_S \int \tau dV \quad cm^{-2}
\end{eqnarray}

\noindent
independent of redshift, for V in units of km~s$^{-1}$.

Direct imaging in HI 21~cm emission is the only reliable method for
determining atomic gas masses. Current efforts have been limited both by
instrumental sensitivity and by accessible frequency coverage to
red-shifts less than about 0.1. The situation is summarized in
Figure~2, where ``Detection'' and ``Imaging'' atomic masses are shown as a
function of red-shift for the WSRT and the SKAI for an integration time
of 100 hours. ``Detection'' has been
defined as requiring a 5$\sigma$ signal in a single 50~km~s$^{-1}$
velocity channel, while ``Imaging'' has been defined as requiring a
5$\sigma$ signal in each of 6 independent 50~km~s$^{-1}$ velocity
channels. The dotted line between red-shifts of 0.2 and 2.5 for the WSRT
indicates the frequency range where receiver systems, while available,
are not yet optimized and are about a factor of 2--3 worse than shown.
The atomic gas masses of the well-known nearby systems M33 and
M101 have been included for reference in the figure, as well as the atomic
gas mass of the ultra-luminous FIR galaxy, III~Zw~35.

As can be seen in
the figure, gas-rich systems
will soon be detectable out to red-shifts of a few tenths with the WSRT.
The SKAI, on the other hand, will allow detection of even low mass
spirals like M33 to z~$>$~1 and gas-rich systems to z~=~3 or more.

Since every narrow velocity interval is so sparsely populated with
condensed atomic gas (at least since the epoch of re-ionization)
observations of this type will not be source confusion limited, even
with only a modest angular resolution of several arcmin. (This same
comment
applies to all emission line tracers of high redshift gas, except
perhaps where the redshift has placed the line frequency near another
emission line of Galactic or terrestrial origin.)

\begin{figure}[htb]
\caption{We show the detection (5$\sigma$ in 50~km~s$^{-1}$) and
imaging (5$\sigma$ in six channels of 50~km~s$^{-1}$) limits of
atomic gas mass as a function of redshift with
current and planned instrumentation.
}
\end{figure}

An important point to note is that the explicit redshift dependence of
the equation for the atomic gas mass in terms of the observed 21~cm line
integral is not often stated. For clarity, we give the equation
below for an optically thin distribution of neutral hydrogen:

\begin{eqnarray}
M_{HI} & = & {16 \pi m_H D_L^2 \over 3 h c A_{21} (1+z)}  \int S_\nu dV \\
       & = & 2.35 \times 10^5 { D_L^2 \over (1 + z) } \int S_\nu dV
\quad M_\odot
\end{eqnarray}

\noindent
for the luminosity distance $D_L$ in Mpc, and the line integral in
Jy~km~s$^{-1}$.

\section{Molecular Gas}

Molecular hydrogen gas is seen directly in the optical band in only
one high redshift QSO absorption-line system 0528-25 at redshift
2.8. In the millimeter band, four objects have now been observed which
show high redshift absorption in various molecules (CO, HCO+, HCN,
O$_2$) generally associated with absorbers in gravitationally lensed
systems (c.f. Combes and Wiklind, these proceedings).  Actual
conditions in proto-galaxies, etc. are not yet clear enough to make
solid predictions, but it is obvious that molecular studies at high
redshift have much to tell us in the near future. With conditions
similar to, say, our Galaxy, gas phase and surface reactions produce
molecular species readily on short time scales $ \sim 10^6 -10^7
yr$. Molecular hydrogen will form rapidly once the density and column
density are high enough.  A thorough discussion of the physical
conditions and the constraints imposed on the $H_2$ species are given
in Black et al (1987). Shielding by dust may be a crucial ingredient
but probably the most important parameter is the strongly evolving
background radiation field.

  The beautiful data on the two lensed objects that
show CO emission at high redshift are well described at this meeting by
Barvainis and Scoville for the Cloverleaf and F10214+4724
respectively.

Frayer (1995) reviews the current evidence for detection of CO emission
at high redshift. Only very tentative detections
have yet been made in non-lensed systems.
When detected at modest redshift (z~$\sim$~0.1), the empirical Galactic
conversion factor suggests molecular gas masses of a few times
$10^{10}M_{\odot}$ concentrated within regions of a few kpc in diameter
(Scoville et al. 1991). However, it has not yet been demonstrated that
a similar conversion factor of CO luminosity to molecular hydrogen mass
need apply under the extreme physical circumstances of circumnuclear
starbursts. Even when
multiple CO line transitions are observed, it is worrisome that they need
not necessarily arise from regions sharing the same physical conditions,
so that the line ratios may remain difficult to interpret.

An impression of the current and future capabilities for imaging
molecular mass at high redshift via associated CO emission is given
for an 8~hour integration in Figure~3. The empirical Galactic
conversion factor (eg. Scoville et al. 1991) gives:

\begin{equation}
 M_{H_2} = 1.2 \times 10^4 { D_L^2 \over (1 + z) } \int S_{CO
1\rightarrow0} dV \quad M_\odot
\end{equation}

\noindent
for the luminosity distance $D_L$ in Mpc, and the CO~1$\rightarrow$0
line integral in Jy~km~s$^{-1}$. If another CO line transition is used
then the constant in eqn.~5 should be scaled in accordance with the
ratio of line luminosities (in erg~s$^{-1}$). We have defined
``detection'' and ``imaging'' as before for the atomic gas mass above
and assumed that the CO~2$\rightarrow$1 transition would be observed
with a 3 times higher line luminosity than that of CO~1$\rightarrow$0.
Nearby normal spirals and two ultra-luminous FIR systems are included
for reference in the figure. The question marks are used to indicate
the uncertainty in assigning a molecular mass to the observed emission
line luminosity in the case of the extreme starburst systems.

{}From Figure~3 it is clear that the MMA (the proposed NRAO Millimeter
Array) should allow study of unlensed
ultra-luminous systems out to redshifts greater than 1, although normal
spirals will only be accessible out to about a tenth.

\begin{figure}[htb]
\caption{
We show the detection (5$\sigma$ in 50~km~s$^{-1}$) and
imaging (5$\sigma$ in six channels of 50~km~s$^{-1}$)
limits of molecular gas mass utilizing the CO~2$\rightarrow$1 transition
as a function of redshift with current and planned instrumentation.
}
\end{figure}

Megamasers are frequently seen in association with relatively edge-on
starburst galaxies and it may well be worth searching systematically
for megamasers at high redshift since their luminosities are so large.
Megamaser emission in the OH, H$_2$CO and H$_2$O lines is a valuable
probe of circumnuclear kinematics with ultra-high angular resolution
and once the theory adequately catches up with the observations it may
also be of help in understanding the physical conditions in extreme
star-bursting systems.

In Figure 4 we show the detection and imaging limits of OH megamaser
emission in a 24 hour integration. Reference luminosities of the sources
in III~Zw~35 and IR20100-4156 are indicated as well as the kilomaser
emission seen in NGC~253. Comparable luminosities to those of OH in
III~Zw~35 are also seen in the H$_2$CO and H$_2$O megamaser sources
(Henkel and Wilson 1990, Baan et al. 1993, Henkel et al. 1984). With
the added frequency coverage of the upgraded WSRT, megamaser emission
should already be detectable to red-shifts greater than about 1,
while the added
sensitivity of the SKAI should allow such sources to be studied in
detail at any redshift.

\begin{figure}[htb]
\caption{
We show the detection (5$\sigma$ in 50~km~s$^{-1}$) and
imaging (5$\sigma$ in six channels of 50~km~s$^{-1}$)
limits for OH mega-maser emission as a function of redshift with
current and planned instrumentation.
}
\end{figure}

There have been reports of ultra cold gas that could constitute a
significant fraction of the dark matter in the Universe ( Lequeux,
Allen and Guilloteau 1993, Pfenniger and Combes 1994, Gerhard and Silk
1995). The absence of such gas in the absorption line spectra of QSOs
indicates that the covering factor of this gas in a sight-line to a
distant QSO is less than $\leq 1\%$. This limit may be a severe
constraint on the proposal that such cold gas is a major constituent
of the Universe.

\section{ Dust}

Some time ago, Ostriker and Heisler (1984) proposed that the observed
fall off in QSO number density might be due to obscuration by
dust. The excellent study of Shaver (1995) shows that this is not
the case. More moderate obscuration is probably present giving
variations in the inferred number counts as a function of redshift for
QSOs of less than order unity. This is consistent with calculations
done by Fall and Pei (1994).

The importance of radio surveys cannot be underestimated here since a
complete radio survey can be used independent of the dust obscuration
and as noted by Shaver (1995) quasars at redshift $z~=~6$ can be easily
seen once the target radio source is known.

Submillimetre observations at high redshift (Isaak et al 1994, McMahon
et al 1994) show that dust masses at redshifts $z~=~4-5$ of order
$10^8M_{\odot}$ and temperatures of say $60~K$ can already be detected.
Protogalaxies may have such dust masses after an initial burst of star
formation and a more or less immediate ( $10^6-10^7 yr$) giant and
supergiant dust producing phase. Conversion of observed continuum
luminosities to actual dust masses remains very tricky while the
emission spectrum is only poorly sampled and there may well be multiple
temperature components present.

In Figure 5 we illustrate the possibility for detecting dust
continuum emission via an 8 hour observation at 230~GHz with the heterodyne
receiver system of the JCMT and the proposed MMA. Dust emission spectra
were calculated with a modified black body ($\nu^{1.5}B(T,\nu)$) using
dust temperatures of 60~K (solid curves) and 30~K (dashed curves). The
ratio of rest-frame 100~$\mu$m to 230~GHz flux density in these cases is
1800 and 360 respectively.
``Detection'' is defined as requiring a 5$\sigma$ signal and ``Imaging''
a 30$\sigma$ signal. Reference luminosities of normal spirals and
the ultra-luminous systems III~Zw~35 and B1202-0725 are indicated.
Single (sub-)millimeter dishes will be able to do better than the
indicated heterodyne JCMT performance through the use of high bandwidth
bolometric detectors (such as SCUBA). This detection
method is not yet applicable to coherent, high resolution imaging with an
interferometric array, although the development of ``hot electron
bolometers'' may make this a possibility in the future.

\begin{figure}[htb]
\caption{We show the limiting luminosity for a continuum observation at
230~GHz for detection (5$\sigma$) and imaging (30$\sigma$) of the dust
continuum as a function of redshift for current and planned
instrumentation. The solid curves are for a modified black body
($\nu^{1.5}B(T,\nu)$) using a dust temperature of 60~K, while the
dashed curves are for a dust temperature of 30~K.
}
\end{figure}

{}From the figure it is clear that the dust continuum becomes more
accessible to a 230~GHz observation beyond z~=~1 for dust temperatures
greater than about 30~K. However, current sensitivities will limit
detection to extreme systems, like B1202-0725, and even then these will
be preferentially found at z~$<$~0.3 and perhaps at z~$>$3. The MMA will
allow comparable detections on less extreme systems (with L$_{FIR}$
``only'' $\sim 10^{12}$~L$_\odot$). Dust continuum from normal spirals
like M33 and M101 will still only be accessible in the local universe.

Many authors have pointed out the excellent correlation of dust
continuum emission and non-thermal radio continuum emission based on
the many thousands of nearby galaxies detected with IRAS. Although still
not understood in detail, there appears to be a strong coupling of both
emission tracers to the massive star formation rate. With this in mind,
we have illustrated in Figure~6 the limiting 1.4~GHz luminosities of
current and planned instrumentation as a function of redshift. A power
law emission spectrum of the form $S \propto \nu^{-0.7}$ has been
assumed. Detection and imaging of luminous star-bursting systems should
already be possible with the VLA out to z~$>$~1. The greater sensitivity
of the SKAI will allow even normal spiral galaxies to be visible out to
cosmological distances.

In this case, of observing the faint continuum emission from distant
sources, it is critical that enough angular resolution be employed so
that source confusion does not limit the sensitivity of an observation.
The deepest existing radio continuum observations, as well as
experience with the HST, suggest that angular resolutions of 0.1--1
arcsec are sufficient to completely circumvent the problem of source
confusion. It is for this reason that the curves in Figure~6 have been
drawn for the VLA (the NRAO Very Large Array) and SKAI, where such
angular resolutions will be achievable, rather than for the WSRT, for
which continuum source confusion at faint flux levels will be a
limitation.

\begin{figure}[htb]
\caption{
We show the limiting luminosity for a continuum observation at
1.4~GHz for detection (5$\sigma$) and imaging (30$\sigma$) of the
non-thermal continuum associated with massive star formation
as a function of redshift for current and planned
instrumentation. The curves assume a power law spectrum of the form
$S \propto \nu^{-0.7}$.
}
\end{figure}

\section{ Cosmology: The Cool Gas History of the Universe}

Observational tests for the detection of the cool pre-ionization ($z
\geq 5$) IGM have been considered by Scott and Rees (1990, also see
Kumar et al. 1995). If the hydrogen spin temperature is greater than
the CMB temperature (T$_R$) at these epochs an emission signature from
neutral hydrogen would be expected. Proto-cluster mass enhancements are
likely to have total masses of 10$^{15}$~M$_\odot$ on proper scales of
less than about 3~Mpc, corresponding to less than about 15~arcmin at
z~=~6. The instrument best-matched to this problem would have a
comparable beam size of some 15~arcmin at an observing frequency of
200~MHz. The necessary telescope diameter of some 350~m corresponds
roughly to that of the individual elements of the SKAI.

An observing
mode that is being envisioned for SKAI is one whereby the
auto-correlations of the individual elements are incoherently summed to
give a $\sqrt N$ increased sensitivity over an individual element,
which still falls short by $\sqrt N$ from the sensitivity of the
coherently combined data, but has a factor of about 10$^4$ greater
brightness sensitivity.
In this mode atomic gas masses of 6$\times
10^{11}$~M$_\odot$ could still be detected at z~=~6. As long as the mass
fraction of neutral atomic hydrogen is greater than about 6$\times
10^{-4}$ then proto-cluster enhancements should be seen. In the case of
the GMRT (the Giant Meterwave Radio Telescope),
the limiting neutral atomic fraction for proto-cluster
detection is about 0.025 (Kumar et al. 1995).

An interesting alternate possibility is that the HI may be observable in
absorption, particularly for high baryon density Universes where the
effect of collisions can drive down the spin temperature, T$_S$, below the
cosmic background radiation temperature, T$_R$.
For a clumpy gas distribution at high
redshift the resulting spin temperature and column density variation
can produce a patchy structure across the sky. Since the HI brightness
temperature is given by:

\begin{equation}
T_B = T_S (1 - e^{-\tau}) - T_R  (1 - e^{-\tau})
\end{equation}

\noindent
the absorption detection signature will be a factor of $T_R/T_S$
stronger than that in emission. In the most extreme scenarios, Scott and
Rees predict $T_R/T_S = 10$, resulting in an easily detectable signal
for instruments like SKAI and possibly the GMRT operating near 200~MHz.

Generally, massive structures are needed to produce a currently
observable effect. However from the work of Steinmetz (1995) and
Kauffman (these proceedings) it is clear that in the standard bottom
up scenarios there are not many really big lumps of neutral gas at high
redshift but there are many small lumps clustering up to large scales
but only at the present epoch. A particularly interesting way to view
this is with the tree diagrams in Lacey and Cole (1994) that indicate
how the dark matter halos put themselves together hierarchically to
form larger galaxies.

 \section{ Active Galaxies, Radio Galaxies and Quasars}

Although it has proved exceedingly difficult to detect the extended
gaseous halo structure around protogalaxies (Djorgovskii et al 1995)
it has been far more productive to look around active galaxies. Large
masses of ionized and cool gas have been found around high redshift
galaxies (c.f. McCarthy 1989). Similarly interesting results
can be found in R\"ottgering et al (1995) and Van Ojik (1995).

 Very recently, however, detections of luminous Lyman Alpha emission
from  protogalaxies at high redshift are emerging from detailed studies
using HST and Keck (Giavalisco et al 1995, Moller et al 1995).
Typically, we might expect the masses of HI associated with the Lyman
Alpha  emitter to be of order $10^{10} M_{\odot}$ although this depends
on a number of  uncertain parameters such as the ionization balance,
etc. Figure~2 suggests that such atomic gas masses should be detectable
with SKAI out to z~$\sim$~3.

 \section{Galaxy Formation: Can it be Observed as Cool Gas?}

 Interestingly, there now seems to be a consensus building about the
pattern of galaxy evolution from combined HST and ground based (CFHT,
Keck) data (Lilly et al. 1995, Driver et al 1995, Griffiths et al
1995). Massive galaxies do not seem to be evolving whereas smaller
dwarf irregulars seem to burst into life between a redshift of $z= 0.5
- 1$ and then fade away by the current epoch. This pushes back the
epoch of massive galaxy formation to redshifts of order $z \geq 3$.

\begin{figure}[htb]
\caption{Simulated spectra of the low mass spiral galaxy M33 are shown
for frequencies between about 10$^8$ and 10$^{14}$~Hz after being
red-shifted to z~=~0.25, 1 and 4, under the assumption of no spectral
evolution. Instrumental sensitivities (1$\sigma$) of existing and planned
instruments are overlaid for both spectral line observations (top panel)
and continuum observations (bottom panel). Spectral line IDs for some of
the major emission lines are indicated at z~=~0.
}
\end{figure}

We note in passing that, apart from the standard collapse and infall
of HI, there are large masses of hot gas such as those associated with
cooling flows in clusters that have cooler denser material at their
centers.  In fact large HI masses are inferred from shadowing effects
(Allen and Fabian 1994).  What this might be like at higher redshift
has been discussed recently by Nulsen and Fabian (1995). Interesting
limits on the cold gas content of the intracluster medium for nearby
clusters of galaxies indicate that the total cold neutral gas content
in the central regions of such clusters is $ \leq 10^9 M_{\odot}$
(O'Dea et al 1995).

Current theories of galaxy formation (c.f. Navarro, Frenk and White
1995) indicate that typical galaxy masses increase as a function of
redshift from dwarf galaxy sized objects at redshifts of order a few
to more massive galaxies at redshift of order unity to cluster sized
objects at the current epoch. We next illustrate how our observational
capabilities overlay the redshifted spectral energy distributions of
several galaxy types and masses.

\begin{figure}[htb]
\caption{Simulated spectra of the massive spiral galaxy M101 are shown
for frequencies between about 10$^8$ and 10$^{14}$~Hz after being
red-shifted to z~=~0.25, 1 and 4, under the assumption of no spectral
evolution. Instrumental sensitivities (1$\sigma$) of existing and
planned instruments are overlaid for both spectral line observations
(top panel) and continuum observations (bottom panel). Spectral line
IDs for some of the major emission lines are indicated at z~=~0.
}
\end{figure}

In Figure 7 we show a simulated spectral energy distribution of the low
mass spiral galaxy M33 redshifted to z~=~0.25, 1 and 4 assuming no
spectral evolution. In the top panel we have overlaid the 1$\sigma$
sensitivity at a spectral resolution of 10$^4$ of a variety of existing
and planned instruments on these spectra. Comparison of the instrument
sensitivities with emission line intensities in the spectra illustrates
out to what redshift such an object might be studied. In the lower panel
the same spectra are overlaid with 1$\sigma$ continuum sensitivities of
the same instruments. In this case the sensitivities should be compared
with the flux densities of the continua to assess out to what redshift
the object might be studied. Integration times of ``one transit'' were
assumed which were typically 8 hours for ground-based telescopes and
10$^4$ seconds for satellite observatories. The various line and
continuum emission components in the model spectra are described in
detail in Braun (1992). A similar set of redshifted spectra and
overlaid instrumental capabilities are shown for the luminous spiral
galaxy M101 in Figure~8. In Figure 9 we show the same plots for the
ultraluminous FIR starburst galaxy III~Zw~35 including its observed
megamaser emission in OH and H$_2$CO.

\begin{figure}[htb]
\caption{Simulated spectra of the luminous starburst galaxy III~Zw~35
are shown for frequencies between about 10$^8$ and 10$^{14}$~Hz after
being red-shifted to z~=~0.25, 1 and 4, under the assumption of no
spectral evolution. Instrumental sensitivities (1$\sigma$) of existing
and planned instruments are overlaid for both spectral line
observations (top panel) and continuum observations (bottom panel).
Spectral line IDs for some of the major emission lines are indicated at
z~=~0.
}
\end{figure}

Comparison of the redshifted model spectra with our current and
projected observational capabilities (in Figs.~7--9) gives us grounds
for guarded optimism about our prospects for studying the galaxy
formation process. Near z~=~1 we should be able to give a very good
characterization of the types of objects which have formed via their
atomic masses and the luminosities of the molecular, dust and stellar
components. The more massive and luminous end of the distribution can
be tracked all the way out to z~$>$~4, while even the low mass end of
the distribution should yield its secrets out to z~$\sim$~0.5. New
instrumentation will be critical to realizing this goal. The
unprecedented sensitivity of the VLT and Keck will be necessary to
permit optical and near-IR spectroscopy to identify these distant
systems. Similarly, the next generation of cm/dm and mm/sub-mm arrays
(SKAI and the MMA) will be needed to ascertain the associated cool
gaseous masses and its kinematics. And although ISO makes an important
contribution to the intervening frequency interval, it is clear that a
new mission with SIRTF (or better) sensitivity will be needed to
effectively fill in the mid-IR to FIR gap.

Great progress is being made in studying the Universe at high redshift
at present by work done with Keck and HST.  After completing this
paper and contemplating the results of the simulations it is clear
that extraordinary progress can be made with the planned
instrumentation. It is obvious how the proposed studies at longer
wavelengths from low frequency radio to sub-millimeter can give vital
information in our quest to understand the physics of the Universe at
high redshift when it was a fraction of its current age.

We thank many of our colleagues for stimulating discussions of this
interesting subject and, in particular, F. Briggs, G. de Bruyn,
R. Ellis, A. Fabian, and M. Rees.

\end{document}